\documentclass[onecolumn,nofootinbib,amsmath,prd,aps,superscriptaddress,tightenlines,preprintnumbers]{revtex4}
\pdfoutput=1

\usepackage{amsmath}
\usepackage{amssymb}
\usepackage{graphicx}
\usepackage{color}
\usepackage{tikz}
\RequirePackage[normalem]{ulem} %DIF PREAMBLE
\RequirePackage{color}\definecolor{RED}{rgb}{1,0,0}\definecolor{BLUE}{rgb}{0,0,1} %DIF PREAMBLE
\usepackage[utf8]{inputenc}
\usepackage{amsmath}
\usepackage{amsfonts}
\usepackage{amssymb}
\usepackage{physics, slashed}
\usepackage{graphicx}
\usepackage{lmodern}
\usepackage{xcolor}
\usepackage{fancyvrb}
\usepackage{siunitx}
\usepackage{hyperref}
\hypersetup{
	colorlinks=true,
	linkcolor=blue,
	filecolor=magenta,      
	urlcolor=cyan,
	pdftitle={Overleaf Example},
	pdfpagemode=FullScreen,
}
\usepackage{cleveref}

\begin{document}

\newcount\hour \newcount\minute
\hour=\time \divide \hour by 60
\minute=\time
\count99=\hour \multiply \count99 by -60 \advance \minute by \count99
\newcommand{\mydate}{\ \today \ - \number\hour :00}
\newcommand{\andre}[1]{\textbf{\color{red} #1}}
\newcommand{\juan}[1]{\textbf{\color{blue} #1}}
\title{Introducing \texttt{NatPy}, a simple and convenient Python module for dealing with  natural units. }
\newcommand{\natpy}{\texttt{NatPy}}
\def\coepp{ARC Centre of Excellence for Dark Matter Particle Physics, Department of
	Physics, University of Adelaide, Adelaide, South Australia 5005, Australia}

\author{Tomas L. Howson}
\email{tomas.howson@adelaide.edu.au}

\affiliation{Special Research Centre for the Subatomic Structure of Matter (CSSM), Department of Physics, University of Adelaide, Adelaide, South Australia 5005, Australia}

\author{Andre Scaffidi}
\email{andre-joshua.scaffidi@to.infn.it}

\affiliation{Istituto Nazionale di Fisica Nucleare, Sezione di Torino, via P. Giuria 1, I–10125 Torino, Italy}

\preprint{ADP-21-10/T1157}

\begin{abstract}
	In high energy physics, the standard convention for expressing physical quantities is natural units. The standard paradigm sets $c = \hbar = \epsilon_0 = 1$ and hence implicitly rescales all physical quantities that depend on unit derivatives of these quantities. We introduce \natpy{}, a simple Python module that defines user-friendly unit objects that can be used and converted within any predefined system of natural units. In this note, we will first introduce, then overview, the algebraic methods utilised by the \natpy{} module.
\end{abstract}

\maketitle

\section{Introduction} \label{sec:intro}
In high energy physics, the common practice is to express quantities in a system where
a basis of
physical constants ($c$, $\hbar$, ...) are defined to be dimensionless with a value of 1, a system referred to as \textit{natural units}.
This practice allows for quantities that would otherwise be dimensionally incompatible
in the SI, such as length and time, or mass, momentum, and energy, to be expressed in the same
units and treated as dimensionally equivalent. While such a system greatly eases the complexity of physical calculations, trouble can
occur when attempting to convert quantities from a system such as the SI to a system of
natural units, or vice versa. One has to determine the correct combination of factors of each
of the unit constants, a process that is both tedious and prone to error. To address
this issue, we introduce \natpy{}, a Python package capable of determining the correct
conversion to and from a defined system of natural units.

\begin{center}
\begin{figure}[ht]
\centering
\begin{BVerbatim}
>> import natpy as nat
>> P = 1 * nat.MeV * nat.fm**(-3)
>> P.convert(nat.Pa)
<Quantity 1.60217663e+32 Pa>
\end{BVerbatim}
\caption{Converting \SI{1}{MeV.fm^{-3}} to pascals using
\natpy{}, two units that are dimensionally equivalent when $c=\hbar=1$.  }
\end{figure}
\end{center}

\natpy{} leverages the \texttt{astropy.units.core.Unit} and \texttt{astropy.units.quantity.Quantity} classes from the \texttt{astropy} package \cite{refId0,2018} in order to allow users to define syntax friendly unit objects for seamless integration into any Python workflow. The power of \natpy{} manifests when large and/or complicated expressions are functions
of many dimensional quantities.  In \cref{section::Intall} we summarise the installation and usage of \natpy{}. In \cref{section::conversion_methos} we  outline a generic method for determining the necessary combination of factors of the unit constants for a conversion before finally discussing the implementation of this method in \cref{sections::algebra_imp,sections::tech_imp}.

\section{Installation and Usage}
\label{section::Intall}
\natpy{} requires $\texttt{python}\geq3.7$ and  can be installed via pip:
\begin{center}
\begin{BVerbatim}
pip install natpy
\end{BVerbatim}
\end{center}
For instructions on usage, readers can refer to a presentation given at PyHEP 2021, which includes a Binder tag for an interactive tutorial \url{https://github.com/AndreScaffidi/Natpy_pyhep_2021}, as well as the package repository  \url{https://github.com/AndreScaffidi/NatPy} and PyPI page \url{https://pypi.org/project/NatPy/}.
%%%%%%%%%%%%%%%%%%%%%%%%%%%%%%%%%%%%%%%%%%%%%%%%%%%%%%%%%%%%%%%%%
\section{Conversion}
\label{section::conversion_methos}

This method aims to determine the necessary factor required to convert a physical
quantity from one set of units to another. This assumes of the use of a generalised natural unit convention
(e.g. $\hbar = c = ... = 1$), and includes determining the necessary combination of
factors of these natural units required to obtain desired units. The outlined method
provides a rigorous 2--step process to determine a conversion factor, without the
trouble of remembering the correct non--natural unit versions of physical quantities. For example, converting energy to mass as $\si{eV}\to \si{eV}/ c^{2}$.

\subsection{Notation}
The quantity $[q]$ with square brackets is defined as the units of the physical quantity $q$. This includes exact units maintaining metric prefixes and other normalisations. So for example
\begin{equation}
	m_e = \SI{512}{keV} \implies [m_e]=[\si{keV}]
\end{equation}
Second, the quantity $\{q\}$ with curly brackets is the dimensions of the quantity $q$. In the same example \begin{equation}
	m_e = \SI{512}{keV} \implies \{m_e\} = \{\mathrm{mass}\}.
\end{equation}
Importantly, dimensions removes any overall factors ($\{\mathrm{years}\} = \{\mathrm{seconds}\}=\{\mathrm{time}\}$), but $\{\mathrm{length}\}\neq \{\mathrm{time}\}$ regardless of natural units.
Finally, we have $\abs{q}$ as the value of the quantity in units of $[q]$ (not absolute value), so \begin{equation}
	m_e = \SI{512}{keV} \implies \abs{m_e}=512.
\end{equation}
As a result, a quantity may be decomposed as \begin{equation}
	q = \abs{q}[q]
\end{equation}
\subsection{Method}
Our method from here outlines a two-step process to determine the conversion
factor from one set of units to another. The process begins with extracting the initial
units, dropping the value of quantity and maintaining the precise units, including
overall multipliers. Consider the example quantity conversion: \begin{equation*}
	\SI{1}{MeV}\qq{to}\SI{5.07d+3}{nm^{-1}}
\end{equation*}
This quantity is given in units of megaelectronvolts (\si{MeV}). We
wish to display this quantity in units of inverse nanometres (\si{nm^{-1}}). Such a conversion requires determining a relevant factor of $c$ and $\hbar$ as well as an overall scalar factor, which can
be troublesome to determine. We let $\abs{q}_i$, $[q]_i$ be the pre--conversion value and units of $q$, and $\abs{q}_f$, $[q]_f$ the post--conversion.
A conversion aims to solve the expression \begin{equation}
	\abs{q}_f[q_f]=\abs{q}_i[q]_i.
	\label{constant}
\end{equation}
The method aims to find a conversion factor $x$ such that  \begin{equation}
	\abs{q}_f = x\cdot \abs{q}_i,
\end{equation}
by solving \begin{equation}
	[q]_f = x^{-1}\cdot [q]_i,
\end{equation}
The factor $x$ has two contributions; one from the overall conversion factor between
``same dimensional'' quantities, denoted $f$ (e.g. metres to centimetres, seconds to
years), and a factor due to the ``natural dimensions" denoted $d$ (e.g. metres to
seconds, mass to energy). The first step in this method is to determine the ``natural dimensionality" of our conversion, i.e. the combination of factors of basis physical constants ($\hbar$, $c$, etc.) necessary to achieve our desired conversion. We determine the necessary product of natural units that has the same dimension of $\{q\}_f/\{q\}_i$, so \begin{equation}
	d=\hbar^{n_{\hbar}}c^{n_c}...,\qq{such that} \{d\}=\{q\}_f/\{q\}_i.
\end{equation}
Hence we must first obtain $d$ from $\{q\}_f/\{q\}_i$, which in our example is \begin{gather*}
	\{d\}=\frac{\{\si{nm^{-1}}\}}{\{\si{MeV}\}}\\
	=\frac{1}{\{\mathrm{length}\}\{\mathrm{energy}\}} = \frac{1}{\{\mathrm{energy}\}\{\mathrm{time}\}}\frac{\{\mathrm{time}\}}{\{\mathrm{length}\}}=\{\hbar\}^{-1}\{c\}^{-1}\\
	\implies d = (\hbar c)^{-1}
\end{gather*}
From here we carefully consider $d$, as this can be expressed in a variety of units. We have from the convention of natural units, that $d=\abs{d}[d]=1$ for any $d$ composed of basis physical constants. Due to this, equation (\ref{constant}) can be written \begin{equation}
	\abs{q}_f [q]_f = \abs{d}[d]\cdot\abs{q}_i [q]_i,
\end{equation}
giving our conversion equations \begin{equation}
	\abs{q}_f = f \abs{d}\abs{q}_i,
\end{equation}
and
\begin{equation}
	\label{finv}
	[q]_f = f^{-1}[d][q]_i,
\end{equation}
where $\abs{d}$ corresponds to the value of $d$ in units of $[d]$.
The second step from here is to determine the multiplicative factor $f$ to match the remaining units. We have by construction of $d$ that $[q]_f$ and $[d]\cdot [q]_i$ have the same dimensions, so any difference in units is simply a scalar multiplicative factor. So $f$ is determined as in equation (\ref{finv}).
In our original example, if $d=(\hbar c)^{-1} = \SI{5.07}{eV^{-1}.\micro m^{-1}}$,\begin{gather*}
	f = [q]_f^{-1} [d][q]_i \\
	= ([\si{nm}]) ([\si{eV}]\cdot [\si{\micro m}])^{-1}([\si{MeV}])\\
	=\frac{[\si{nm}]}{[\si{\micro m}]}\frac{[\si{MeV}]}{[\si{eV}]}
	= \num{d-3}\times\num{d+6}=\num{d+3}
\end{gather*}

Finally, we have that $x = \overline{d}\cdot f$, which is our final conversion factor. In our example,
\begin{gather*}
	x = \num{5.07d+3}.
\end{gather*}
So as a result,
\begin{gather*}
	\overline{q}_f = \num{5.07d+3} \overline{q}_i,\\
	\mathrm{where}\quad[q]_i=\si{MeV},\qq{and}
	[q]_f = \si{nm^{-1}},\\
	\implies \SI{1}{MeV}=\SI{5.07d+3}{nm^{-1}}
\end{gather*}

To summarise the method;
\begin{enumerate}
	\item Find the dimensions of the quotient of the final and initial units, $\{d\}=\{q\}_f/\{q\}_i$,
	\item Determine the combination of factors that has the same dimensions as that
	      quotient, $d=\hbar^{n_{\hbar}}c^{n_{c}}...$,
	\item Find the resulting scaling factor from cancelling any remaining units,
	      $f=[q]^{-1}_f [d][q]_i$,
	\item Obtain the final conversion factor, $x=\abs{d}\cdot f$.
\end{enumerate}

\section{Algebraic Implementation}
\label{sections::algebra_imp}
The method expressed above is still somewhat tedious. While it does give a systematic
framework from which to find conversion factors, step 2 in which $d$ is determined
from the dimensionality is not automatic, requiring somewhat arbitrary algebraic manipulations until
the correct factor is found. Instead, we propose a direct computation to determine
the factor $d$ using simple linear algebra. It is this implementation on which \natpy{} is
developed, applying this computation to dimensional quantities to obtain conversion factors.

\subsection{Notation}
In this algebraic framework, we shall define two sets of vectors; The powers of
dimensionality (PoD) of a quantity $q$, denoted $\widetilde{q}$, and the powers of natural
units (PoNU) of a dimensional conversion factor $d$, denoted $\vec{d}$. The PoD of a quantity $q$ is the vector of multiplicities of
the dimensions of $q$ in terms of a set of base units. The seven SI base units are used in
\natpy{}. \begin{equation}
	\{q\}=\{\mathrm{length}\}^{n_{\mathrm{length}}}\{\mathrm{time}\}^{n_{\mathrm{time}}}\{\mathrm{mass}\}^{n_{\mathrm{mass}}}...
	\implies \widetilde{q}:=\begin{pmatrix}
		n_{\mathrm{length}} \\n_{\mathrm{time}}\\n_{\mathrm{mass}}\\\vdots
	\end{pmatrix}.
\end{equation}
For example, a force $F$ has a PoD given by: \begin{gather*}
	\{F\} =
	\{\mathrm{newton}\}=\frac{\{\mathrm{mass}\}\{\mathrm{length}\}}{\{\mathrm{time}\}^2}\\
	\implies \widetilde{F}=\begin{pmatrix}
		F_{\mathrm{length}} \\F_{\mathrm{time}}\\F_{\mathrm{mass}}
	\end{pmatrix}=\begin{pmatrix}
		1 \\-2\\1
	\end{pmatrix}.
\end{gather*}
The powers of natural units (PoNU) of the dimensional conversion factor $d$ is defined
similarly, but by the powers of the basis physical constants rather than base units. So \begin{equation}
	d=\hbar^{n_{\hbar}}c^{n_{c}}... \implies \vec{d}:=\begin{pmatrix}
		n_{\hbar} \\n_{c}\\\vdots
	\end{pmatrix}.
\end{equation}
\subsection{Implementation}
Given the previously defined notation, the goal of the outlined method is
simple; find $\vec{d}$ given $\widetilde{d}$. Since $d$ is defined entirely by $\vec{d}$, the PoNU is now the target quantity to calculate. The dimensionality factor $d$ is found such that $\{d\} = \{q\}_f/\{q\}_i.$ Due to the
property that $a^ba^c = a^{b+c}$, it follows that $\widetilde{d}=\widetilde{q}_f -
	\widetilde{q}_i$, giving a simple method to determine $\widetilde{d}$. Furthermore,
applying this same property to the definition of $d=\hbar^{n_{\hbar}}c^{n_{c}}...$, we
have that \begin{gather*}
	\begin{pmatrix}
		d_{\mathrm{length}} \\
		d_{\mathrm{time}}   \\
		d_{\mathrm{mass}}   \\
		\vdots
	\end{pmatrix} = n_{\hbar}\begin{pmatrix}
		\hbar_{\mathrm{length}} \\
		\hbar_{\mathrm{time}}   \\
		\hbar_{\mathrm{mass}}   \\
		\vdots
	\end{pmatrix} + n_c\begin{pmatrix}
		c_{\mathrm{length}} \\
		c_{\mathrm{time}}   \\
		c_{\mathrm{mass}}   \\
		\vdots
	\end{pmatrix} + ...\\
	\widetilde{d} =\begin{pmatrix}
		\hbar_{\mathrm{length}} & c_{\mathrm{length}} & ...    \\
		\hbar_{\mathrm{time}}   & c_{\mathrm{time}}   & ...    \\
		\hbar_{\mathrm{mass}}   & c_{\mathrm{mass}}   & ...    \\
		\vdots                  & \vdots              & \ddots
	\end{pmatrix}\begin{pmatrix}
		n_{\hbar} \\n_c\\\vdots
	\end{pmatrix} = A\vec{d}\\
	\implies \vec{d}=A^+\widetilde{d}
\end{gather*}

We shall define $A$ the change of dimensionality (CoD) matrix.  The final result depends on calculating the left pseudo--inverse
of $A$, denoted $A^+$, obtainable as the Moore--Penrose inverse of $A$. Notice the CoD matrix is
entirely independent of the initial or final units, being a function only of the chosen
system of natural units. This results in $A^+$ needing only to be calculated once, and
is usable for all natural dimensional conversions in a given system of natural units.

The goal of determining $\vec{d}$ from $\widetilde{d}$ has become one of simple algebra;
\begin{enumerate}
	\item The CoD matrix $A$ is constructed from the PoD vectors of each of the basis physical constants,
	      ($\widetilde{\hbar}$, $\widetilde{c}$,...),
	\item The Moore--Penrose pseudo--inverse $A^+$ of $A$ is calculated,
	\item Apply $A^+$ to the PoD vector of the dimensionality factor $\widetilde{d}$
	      to obtain the the PoNU vector of this conversion, $\vec{d}$,
	\item Calculate $d$ from $\vec{d}$.
\end{enumerate}

\section{Technical Implementation}
\label{sections::tech_imp}

\natpy{} implements the above conversion process by leveraging the \texttt{astropy} Python module. It makes use of the \texttt{astropy.units.core.Unit} and \texttt{astropy.units.quantity.Quantity} classes to define dimensional objects. It then draws on the \texttt{astropy.constants} submodule to define the list of constants on which to form the natural unit basis. From this list the change of dimensionality (CoD) matrix is constructed, and pseudo--inverted by the \texttt{numpy.linalg} submodule \cite{numpy}. The \texttt{natpy.convert} method is constructed to calculate the power of dimensionality (PoD) vector between initial and target units, so as to find the power of natural units (PoNU) vector of the dimensional conversion factor $d$. The correct factors of natural units are multiplied to the initial quantity, and \texttt{astropy} is used for the final scalar conversion to the target units.

By utilising the \texttt{astropy} module, \texttt{numpy} is fully incorporated. \texttt{Quantity} objects may store \texttt{numpy.ndarray} objects as the quantity, providing full access to \texttt{numpy}'s \texttt{ufunc} functionality. In storing an array of quantities with the same units, such an array may be converted between compatible units by only calculating a conversion factor once, applying the same factor across the array.

\section{Summary}
\natpy{} provides a computational framework for calculations involving dimensional quantities in a way that properly considers equivalences due to the conventions of a given natural units scheme. Quantities may be converted from the conventional units of high energy physics, such as masses in \si{GeV} or times in \si{fm}, to the standard SI units \si{kg} and \si{s} and vice versa, when considering a system of natural units. \natpy{} automates this process, reducing the likelihood of simple algebraic errors. The framework provided can be incorporated into existing analysis, either by using \natpy{} to find relevant conversion factors, or by storing quantities in \texttt{Quantity} objects and using \natpy{} to convert such an object between compatible units. \natpy{} is fully incorporated with \texttt{numpy} to allow for powerful functionality.

\section{Acknowledgements}
TLH acknowledges support from an Australian Government Research Training Program Scholarship. AS acknowledges support from the research grant “The Dark Universe: A Synergic Multimessenger Approach” No. 2017X7X85K funded by MIUR and the project “Theoretical Astroparticle Physics (TAsP)” funded by the INFN. We thank all funding agencies.
\bibliography{natpy}

\end{document}